\documentclass[showpacs,aps,prl,amsmath,amssymb,twocolumn]{revtex4-1}
\usepackage{graphicx}
\usepackage{dcolumn}
\usepackage{bm,color}
\usepackage{txfonts}
\usepackage{amsmath,epsfig}
\usepackage{epstopdf}
\usepackage{times}
\newcommand{\ket}[1]{\left\vert#1\right\rangle}
\newcommand{\bra}[1]{\left\langle#1\right\vert}

\newcommand{\eq}{Eq.~}
\newcommand{\eqs}{Eqs.~}
\newcommand{\fig}{Fig.~}
\newcommand{\figs}{Figs.~}
\newcommand{\cf} {cf.~}
\newcommand{\ug} {\!=\!}
\newcommand{\tens} {\!\otimes\!}
\newcommand{\piu} {\!+\!}
\newcommand{\meno} {\!-\!}
\newcommand{\ie} {i.e.~}

\newcommand{\lhs} {l.h.s.~}
\newcommand{\vs} {vs.~}

\begin{document}

\author{Francesco Ciccarello}
\author{Vittorio Giovannetti}

\affiliation{NEST, Scuola Normale Superiore and Istituto Nanoscienze-CNR, Piazza dei Cavalieri 7, I-56126 Pisa, Italy}
%\pacs{42.50.Pq, 05.60.Gg}

\title{Creating quantum correlations through local non-unitary memoryless channels}
 
\date{\today}
\begin{abstract}
We show that two qubits, initially in a fully classical state, can develop significant quantum correlations as measured by the quantum discord (QD) under {
the action of  a {\it local} memoryless noise (specifically we consider the case of a Markovian 
amplitude-damping channel)}. This is analytically proven after deriving in a compact form the QD for the class of separable states involved in such a process. We provide a picture in the Bloch sphere that unambiguously highlights the physical mechanism behind the effect {regardless of the specific measure of QCs adopted}.
\end{abstract}
\maketitle

\noindent

The existence of states where two or more systems are correlated in a way unattainable in classical physics ranks among the most puzzling and yet distinctive features of quantum mechanics. Such possibility is commonly pictured in terms of an extra amount correlations, usually referred to as quantum correlations (QCs), which a multipartite system can possess in addition to those of a merely classical nature. Until recently, the scrutiny of QCs has been almost ubiquitously intertwined with investigations on entanglement \cite{horo1} and the pivotal role that it plays in the area of quantum information processing \cite{nc}. A breakthrough yet occurred as soon as it was realized \cite{seminal} that while classicality always entails separability the reverse is in general untrue (a state is entangled iff it is non-separable). Entanglement thereby is not the only form in which QCs can occur. Such finding brought about a widened perspective, which is currently prompting a growing number of researchers to advance the field along various lines.  A prominent one is the quest for faithful easy-to-handle indicators of QCs \cite{measures,modi}. Among those proposed so far, quantum discord (QD) \cite{seminal} is having a considerable impact despite its explicit calculation is usually demanding even for two qubits (\ie a pair of two-dimensional systems). Yet, evidence of its ability to capture QCs not detected by entanglement has been supplied in various frameworks such as one-qubit quantum computation (even experimentally) \cite{datta} and quantum phase transitions \cite{qpt}. 

Another major concern that soon arose is to assess how QCs, according to such a novel paradigm, are affected by non-unitary dynamics. These typically stem from the interaction with an environment, a process where entanglement is extremely fragile in most cases \cite{horo1,nc}. In contrast, QD was proven to be in general quite resilient to such dynamics and, strikingly, in some cases even fully insensitive over long stages \cite{laura}. From a reverse perspective, it was shown that the quantum noise arising from a common bath can create QCs initially fully absent \cite{common}, a phenomenon well-known to occur for entanglement either \cite{common2}. Also, preexisting QCs can exhibit an increasing behavior at some stages of their time evolution in the presence of non-Markovian local channels as recently demonstrated for both entanglement and QD \cite{nonmarkov}. In such instances, the increase of QCs stems either from the ability of a common memoryless reservoir to mediate an effective interaction or as a memory effect of local environments. The question is now raised:  {Can the interaction with a bath which is both {\it local} and {\it memoryless} enhance QCs?} The answer is well-known to be negative for entanglement, which cannot grow under any local quantum maps \cite{horo1}. As for the full amount of QCs, however, the issue is not as much trivial since as stressed above even separable states may feature some quantumness \cite{seminal}. As far as local {\it unitaries} are concerned, though, QCs' measures, such as QD, cannot increase \cite{seminal, measures,modi}. {In some respects, 
this conclusion might be expected to be strengthened with {\it noisy} local operations. { However, QCs without entanglement can arise under a local non-unitary operation owing to its ability to map orthogonal into non-orthogonal states \cite{modi}. } 
{In this work, we present a simple paradigmatic process, where entanglement is absent throughout, which clearly testifies that QCs can even be entirely created (or increased) {\it solely via the interaction with a bath which is both local and memoryless}}. We provide rigorous and comprehensive insight into such effect in a way that makes transparent the underlying physical mechanism. {Although we focus on QD, it will become clear that the essential physical effect takes place regardless of the specific measure of QCs used.}} 

To begin with, we briefly recall the definition of QD. Given two systems $A$ and $B$ in a state $\rho$, this measures the discrepancy between the mutual information $\mathcal{I}$ and the classical correlations $\mathcal{C}$ associated with $\rho$ \cite{seminal}. A local measurement on $B$ in a given orthonormal basis can be specified by a complete set of projectors $\{B_k\}$, where $k$ indexes a possible outcome. If $k$ is recorded with probability $p_k\!=\!{\rm Tr}[B_k\rho B_k]$ the overall system collapses onto the (normalized) state $\rho_k\!=\!(B_k\rho B_k)/p_k$. Then the QD $\mathcal{D}^{\leftarrow}$ can be expressed as \cite{seminal}
\begin{equation}\label{dis}
\mathcal{D}^{\leftarrow}(\rho)\!=\!S(\rho_B)\!-\!S(\rho)\!+\min_{\{B_k\}}\sum_k{p_kS(\rho_k)}\,\,.
\end{equation}
In \eq(\ref{class}), $\rho_{B}\!=\!{\rm Tr}_{A}\rho$ is the reduced density operator that describes the state of $B$, $S(\sigma)\!=\!-{\rm Tr}\,(\rho\,{\rm log}_2 \sigma)$ is the Von Neumann entropy of an arbitrary state $\sigma$ and the infimum in the last term is evaluated over all the possible sets $\{B_k\}$. As for the QD $\mathcal{D}^\rightarrow$ involving measurements on $A$ with associated projectors $\{A_k\}$, this is obtained from (\ref{dis}) through replacement of $B$ with $A$. In general, $\mathcal{D}^\leftarrow\!\neq\!\mathcal{D}^\rightarrow$ \cite{seminal}.
\begin{figure}
 \includegraphics[width=0.45\textwidth]{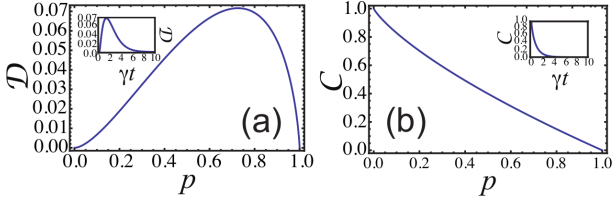}
\caption{(Color online){ Behavior under a local amplitude-damping channel of discord $\mathcal{D}$ (a) and classical correlations $\mathcal{C}$ (b) against $p$. In each panel, the inset shows the corresponding behavior expressed as a function of the rescaled time $\gamma t$, where $p\ug1\meno e^{-\gamma t}$} .
 \label{Fig1}}
\end{figure} 

Specifically, in the process under study $A$ and $B$ are two qubits and the only involved states have the {\it separable} form
\begin{equation}\label{class}
\rho\!=\!\tfrac{1}{2}\left(|0\rangle_A\!\langle0|\!\otimes\!\tau_{0B}+|1\rangle_A\!\langle1|\!\otimes\!\tau_{1B}\right)\,\,,
\end{equation}
where 
{$\tau_{0(1)}$ are generic density matrices while 
 the orthonormal set of vectors}  $\{|0\rangle,|1\rangle\}$ is the usual local computational basis. {The states~(\ref{class})} are commonly dubbed as {\it classical-quantum states} owing to $\mathcal{D}^\rightarrow\ug0$ as is immediate to check \cite{disA}, while in general $\mathcal{D}^\leftarrow\!\neq\!0$. We will thereby set $\mathcal{D}\!\equiv\!\mathcal{D}^\leftarrow$ henceforth.

Consider now the initial state 
\begin{equation}\label{rho0}
\rho_0\!=\!\tfrac{1}{2}\left(|0\rangle_A\!\langle0|\otimes|+\rangle_B\langle+|+|1\rangle_A\!\langle1|\otimes|-\rangle_B\langle-|\right)\,\,,
\end{equation}
\ie in the light of (\ref{class}) $\tau_0\ug|+\rangle\langle+|$ and $\tau_1\ug|-\rangle\langle-|$, where $|\pm\rangle\ug(|0\rangle\!\pm\!|1\rangle)\!/\!\sqrt{2}$. 
Although manifestly correlated, state (\ref{rho0}) is fully classical since it is diagonal in the basis resulting from the tensor product between $\{|0\rangle_A,|1\rangle_A\}$ and $\{|+\rangle_B,|-\rangle_B\}$, namely two local {\it orthogonal} bases \cite{talmor}.  Thus $\mathcal{D}(\rho_0)\ug0$. Assume now that while $A$ is well protected from the external environment so is not $B$, which is in contact with a dissipative Markovian bath. In such a case, the system dynamics as a function of time $t$ is fully described by an amplitude-damping channel \cite{nc}. This has an associated quantum map $\mathcal{E}_p$ that transforms state $\rho_0$ according to $\mathcal{E}_p(\rho_0)\ug E_0 \rho_0 E_0^\dagger\piu E_1 \rho_0 E_1^\dagger$, where $E_0\ug|0\rangle_B\langle0|\piu \sqrt{1\meno p}|1\rangle_B\langle1|$ and $E_1\ug\sqrt{p}|0\rangle_B\langle1|$ are the associated Kraus operators  while $p$ is a probability that grows with time $t$ according to $p\ug1\meno e^{-\gamma t}$ ($\gamma$ is a relaxation rate). As anticipated, $\mathcal{E}_p(\rho_0)$ evidently belongs to class (\ref{class}) for any $p$ (\ie $\forall t$) since $E_0$ and $E_1$ act on $B$ only (hence entanglement never appears throughout). In \fig1(a), we plot $\mathcal{D}$ [\cf \eq(\ref{dis})] against $p$ and the rescaled time $\gamma t$ as resulting from numerical evaluation \cite{numerical}.
$\mathcal{D}(p)$, which is initially null as discussed above, at a first stage {\it grows} and eventually decays to zero as $p\!\rightarrow\!1$, \ie for $t\!\gg\!\gamma^{-1}$. This marks a profound difference between QD and entanglement in that a local non-unitary and memoryless channel is able to create QCs previosly fully absent. Remarkably, the dissipative dynamics is merely detrimental to the classical correlations $\mathcal{C}\ug\mathcal{I}\meno\mathcal{D}$ \cite{seminal}, where $\mathcal{I}\ug S(\rho_A)\piu S(\rho_B)\meno S(\rho)$ is the mutual information \cite{nc}. $\mathcal{C}$ indeed exhibits a monotonic decay vanishing for $p\!\rightarrow\!1$, \ie $\gamma t\!\rightarrow\!\infty$, as shown in \fig1(b).

In the following, we make this result rigorous by analytically deriving $\mathcal{D}(p)$ and $\mathcal{C}(p)$ so as to reproduce \fig1. To this aim, we first explicitly derive in a compact form the QD of states (\ref{class}) when $\tau_0$ and $\tau_1$ have the same purity. Next, we present a picture in terms of trajectories in the Bloch sphere clearly highlighting the physical mechanism that causes QD to necessarily grow in the present process.

Among states (\ref{class}) a prominent instance is the resource state for the B92 quantum cryptography protocol \cite{nc}, which reads 
{$\rho_{{\rm B92}}\!=\!\tfrac{1}{2}\left(|0\rangle_A\!\langle0|\otimes|0\rangle_B\langle0|+|1\rangle_A\!\langle1|\otimes|+\rangle_B\langle+|  \right)$.}
Similarly to the popular BB84 \cite{nc} 
the above is among those quantum protocols where the exploited quantum resource is not entanglement, which is is fully absent. Rather, it harnesses one-way QCs stemming from the non-distinguishability of states $|0\rangle$ and $|+\rangle$ \cite{talmor}. Later on, we will indeed show that 
{$\rho_{{\rm B92}}$} 
possesses the maximum allowed QD within family (\ref{class}) with $\tau_0$ and $\tau_1$ having equal purities. Also, states such as (\ref{class}) can allow for quantum locking (see \cite{ql} and references therein).
To our knowledge, the literature lacks in explicit formulas for the QD of (\ref{class}). We will thus carry out an {\it ab initio} calculation.

To this aim, our first step is to express the single-qubit states $\tau_0$ and $\tau_1$ in \eq(\ref{class}) through the Bloch-sphere representation as $\tau_i\!=\!(\openone\!+\!{\bf s}_i\cdot\!\mbox{\boldmath$\sigma$})/2$, where $\openone$ and $\mbox{\boldmath$\sigma$}\!=\!\{\sigma_1,\sigma_2,\sigma_3\}$ are the usual identity and Pauli operators, respectively, while ${\bf s}_i$ is the Bloch vector corresponding to $\tau_i$ ($i\ug0,1$). Without loss of generality we can assume that $s_{0 x}\!=\!s_{0 y}\!=\!0$, $s_{0z}\!\equiv\!s_0$ and $s_{1y}\!=\!0$, \ie in the Bloch sphere ${\bf s}_0$ and ${\bf s}_1$ lie on the $X\!-\!Z$ plane with ${\bf s }_0$ along the $Z$-axis ($s_i\!=\!|{\bf s}_i|\le1$). Indeed, one can reduce the problem to such a case by applying a suitable single-qubit rotation, which cannot affect the QD like any local {\it unitary} operation  \cite{seminal}. Using this along with $|0\rangle\langle0|\!=\!(\openone\!+\!\sigma_3)/2$ and $|1\rangle\langle1|\!=\!(\openone\!-\!\sigma_3)/2$, state (\ref{class}) can be arranged as (from now on we drop subscripts $A$ and $B$)
\begin{equation}\label{decomp}
\rho\!=\!\tfrac{1}{4}\left[\openone\!\otimes\!\openone\!+\!\openone\!\otimes\!(a_1 \sigma_1\!+\!a_3 \sigma_3)\!+\!\sigma_3\!\otimes\!(b_1 \sigma_1\!+\!b_3 \sigma_3)\right]\,\,
\end{equation}
with
\begin{eqnarray}\label{coeffs}
a_1\ug- b_1\ug \tfrac{s_1\!\sin \varphi}{2}\,,\,\,\,a_3\ug \tfrac{s_{0}\piu s_1\!\cos \varphi}{2}\,,\,\,\,b_3\ug \tfrac{s_{0}\meno s_{1}\cos\varphi}{2}\,,\,\,\,\,\,\,
\end{eqnarray}
where we have carried out the replacements $s_{1x}\ug s_1\!\sin \varphi$, $s_{1z}\ug s_1\cos \varphi$ ($\varphi$ is the angle between ${\bf s}_0$ and ${\bf s}_1$).
\begin{figure}
 \includegraphics[width=0.45\textwidth]{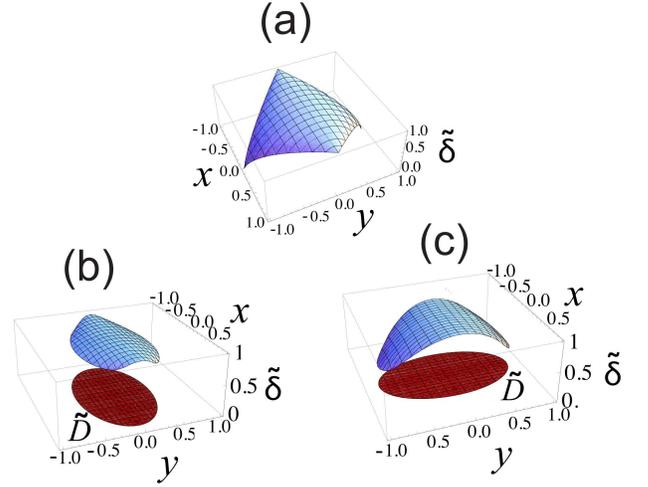}
\caption{(Color online) (a) Function $\tilde{\delta}(x,y)$ for $|x|\!\le\! 1$ and $|y|\!\le\! 1$. (b) and (c) Function $\tilde{\delta}(x,y)$ for $(x,y)\!\in\!\tilde{D}$ as given by \eq(\ref{ellipse}) when $s_0\ug\ug s_1\ug1$ and $\varphi\ug \pi/3$ (b), $\varphi\ug 2\pi/3$ (c). In either case, the red region on the $x\!-\!y$ plane represents the corresponding $\tilde{D}$. \label{Fig2}}
\end{figure}
Next, in the light of (\ref{dis}) we need to calculate how $\rho$ is transformed when a generic Von Neumann measurement is performed on system $B$, \ie $\rho_k$, as well as the associated probability $p_k$. {As in~\cite{luo}} we use the property that the projector corresponding to any such measurement can be expressed as $\openone\otimes B_k$ with $B_k\ug V\Pi_kV^\dagger$, where $\Pi_k\ug\ket{k}\!\bra{k}$ ($k\ug0,1$) and $V$ is a generic one-qubit unitary. We thus expand $p_k\rho_k\!=\!(\openone\otimes B_k) \rho (\openone\otimes B_k)$ as
\begin{eqnarray}\label{rhok}
p_k\rho_k\!
= (\openone\tens V)(\openone\!\otimes\! \Pi_k) (\openone\tens V^\dagger) \rho (\openone\tens V)(\openone\!\otimes\! \Pi_k) (\openone\tens V^\dagger)\,.\,\,\,
\end{eqnarray}
The unitary $V^\dagger$ tranforms each Pauli matrix $\sigma_i$ according to $V^\dagger\!\sigma_i\! V\ug \sum_{j\ug1}^3 \upsilon_{ij}\sigma_j$, where $\upsilon_{ij}$ are real numbers
{satisfying the constraint  $\upsilon_{1j}^2\piu\upsilon_{2j}^2\piu\upsilon_{3j}^2\ug1$, for any $j\ug 1,2,3$. } 
 Also, for $k\ug0,1$ $\Pi_k\sigma_1\Pi_k\ug\Pi_k\sigma_2\Pi_k\ug{\bf0}$ while $\Pi_k\sigma_3\Pi_k\ug f_k \Pi_k$, where $f_0\ug1$ and $f_1\ug-1$. This along with (\ref{decomp}) yield (\ref{rhok}) in the simpler form
\begin{eqnarray} \nonumber 
p_k \rho_k\ug(\!\openone\tens V\!)\!\left[\!\frac{\mu_k\! \openone\piu \nu_k\sigma_3}{4}\tens\Pi_k\!\right]\!(\!\openone\tens V^\dagger\!)\ug\frac{\mu_k \!\openone\piu\nu_k\sigma_3}{4}\tens (V\Pi_k V^\dagger)\,\,,
\end{eqnarray}
where
\begin{eqnarray}  \label{muk}
\mu_k \ug 1\piu f_k(a_1\upsilon_{13}\piu a_3\upsilon_{33})\,, \qquad \nu_k\ug f_k(b_1\upsilon_{13}\piu b_3\upsilon_{33})\,.
\end{eqnarray}
As $V\Pi_k V^\dagger$ represents a pure state and each Pauli matrix is traceless it is immediately checked that $p_k\ug\mu_k/2$. 
Therefore, we obtain that $\rho_k\ug(\!\openone\piu\nu_k/\mu_k\sigma_3)/2\tens (V\Pi_k V)$ whose eigenvalues are $(1\!\pm\!\nu_k/\mu_k)/2$ (each two-fold degenerate). The quantity to minimize entering the last term of (\ref{dis}) thus reads
\begin{eqnarray}\label{delta}
\delta(\upsilon_{13},\upsilon_{33})\ug\sum_{k\ug0,1} p_k S(\rho_k)\ug\sum_{k=0,1} \tfrac{\mu_k}{2} \,h\left(\tfrac{1\!\pm\!\nu_k/\mu_k}{2}\right)\,\,,
\end{eqnarray}
where $h(x)\ug-x\log_2 x\meno(1\meno x)\log_2(1\meno x)$ is the binary Shannon entropy function and on the \lhs we have highlighted the dependence on variables $\upsilon_{13}$ and $\upsilon_{33}$. The identity $\upsilon_{13}^2\piu\upsilon_{23}^2\piu\upsilon_{33}^2\ug1$ (see above) yields that $\{\upsilon_{13},\upsilon_{33}\}$ must fulfill $\upsilon_{13}^2\piu \upsilon_{33}^2\ug 1\meno \upsilon_{23}^2\!\le\!1$, \ie they belong to  the unit circle. 
To work out the infimum of $\delta$ we accomplish the linear transformation $x\ug a_1\upsilon_{13}\piu a_3\upsilon_{33}$ and $y\ug b_1\upsilon_{13}\piu b_3\upsilon_{33}$. This way, (\ref{delta}) now becomes a universal function of $x$ and $y$, which we call $\tilde{\delta}(x,y)$, \ie it no longer depends on parameters $\{a_i, b_i\}$ specifying states $\tau_0$ and $\tau_1$ [\cf \eq(\ref{coeffs})]. 

In \fig2(a) we plot $\tilde{\delta}$ against $x$ and $y$. Note that its actual domain of definition is the square having side length $1\!/\!\sqrt{2}$ and vertices at points ($\pm 1$,0) and (0,$\pm 1$) [this is because of the logarithms of $1\!\pm\!f_k y/(1\piu f_k x)$, see \eqs(\ref{muk})-(\ref{delta})]. Also, $\tilde{\delta}$ is an even function of $x$ ($y$) for any set value of $y$ ($x$) as is also clear from its dependance on $x$ and $y$ through $\mu_k$ and $\nu_k$ [see \eqs(\ref{muk})-(\ref{delta})]. It takes value 1 for $x\ug y\ug0$ and decreases as the distance from the origin grows. Its concavity is minimum along the $x$-axis, where the function is fully flat, and maximum along the $y$-axis [see the sail-like shape in \fig2(a)]. 
As for the domain $\tilde{D}$ within which (\ref{delta}) is to be minimized, \ie the region in the new reference frame corresponding to the unit circle $\upsilon_{13}^2\piu \upsilon_{33}^2\!\le\!1$ (see above), it is immediately checked through the inverse transformation that this is an elliptic region given by
\begin{eqnarray}\label{ellipse}
&\tilde{D}\ug\{ x,y:\,\,\,A\,{x^2}\piu2B\, x y\piu C\,{y^2}\le1\}\;,& \\
\label{ABC}
&A\ug\left(\tfrac{|{\bf s_0}-{\bf s_1}|}{s_0 s_1 \sin \varphi}\right)^2
\,,\,\,B\ug\tfrac{s_0^2-s_1^2}{(s_0 s_1 \sin \varphi)^2}\,,\,\,C\ug\left(\tfrac{|{\bf s_0}+{\bf s_1}|}{s_0 s_1 \sin \varphi}\right)^2&\!.\,\,\,\,\,\,\,
\end{eqnarray}
As is evident from \fig2(a), owing to the concavity of $\tilde{\delta}$ its infimum under constraint (\ref{ellipse}) necessarily lies on the boundary of $\tilde{D}$, \ie the ellipse obtained from (\ref{ellipse}) by turning the inequality into an identity. While such ellipse is centered at $x\ug y\ug0$, its own axes in general do not coincide with those defining the reference frame. 
Yet, in the case that $B\ug0$ [\cf \eqs(\ref{ellipse}) and (\ref{ABC})] the ellipse is not rotated with respect to the $x-$ and $y-$axis. This circumstance
physically occurs when {\it $\tau_0$ and $\tau_1$ have the same purity}, \ie $s_0\ug s_1$ [see \eq(\ref{ABC})]. Henceforth, we will focus on such a case, which is enough for the scopes of the present work and allows for a prompt analytical derivation of the infimum of (\ref{delta}) under constraint (\ref{ellipse}). Hence, by setting $s_0\!=\!s_1\!=\!s$ coefficients (\ref{ABC}) become
$A\ug {1}/{r_x^2}$, $B\ug0$, $C\ug {1}/{r_y^2}$,  
 where $r_x\ug s|\sin(\varphi/2)|$ and $r_y\ug s|\cos(\varphi/2)|$ are the ellipse semi-axis lengths along the $x$- and $y$-axis, respectively.
\begin{figure}
 \includegraphics[width=0.45\textwidth]{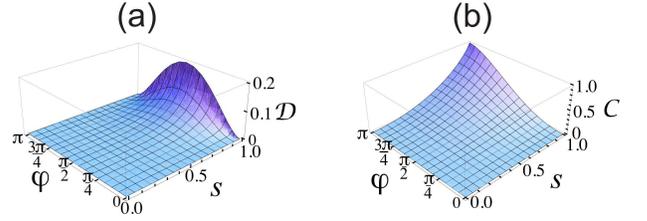}
\caption{(Color online) $\mathcal{D}$ (a) and $\mathcal{C}$ (b) \vs $s$ and $\varphi$. \label{Fig3}}
\end{figure}
Now, because of the shape of $\tilde{\delta}(x,y)$ [see \fig2(a)] when the main axes and the ellipse own ones are collinear $\tilde{\delta}(x,y)$ takes its minimum at the points $(0, \pm r_y)$. This is evident from \figs2(b) and (c), where we plot the restriction of $\tilde{\delta}$ to region $\tilde{D}$ in a paradigmatic case such that $r_x\!>\!r_y$ (b) and one featuring $r_x\!<\!r_y$ (c). The minimum of (\ref{delta}) within region (\ref{ellipse}), \ie  the last term in \eq(\ref{dis}), thus reads
\begin{eqnarray}\label{inf}
\! \min_{\tilde{D}} \tilde{\delta}(x,y)\ug \tilde{\delta}(0,r_y) \ug\tilde{\delta}\left(0, s \left|\cos \tfrac{\varphi}{2}\right|\right)\ug h\left[\tfrac{1\piu s|\sin \varphi/2|}{2}\right]\,,\,\,\,\,\,\,\,\,
\end{eqnarray}
where we used that $\mu_k\ug1\piu f_k x$ and $\nu_k\ug f_k y$ [\cf \eqs(\ref{muk})-(\ref{delta})]. 

As for $S(\rho_B)$, the trace over $A$ of (\ref{decomp}) is obtained as $\rho_B\ug (\openone\piu a_1 \sigma_1\!+\!a_3 \sigma_3)/2$, whose eigenvalues with the help of (\ref{coeffs}) are found as $(1\!\pm\! s |\cos\varphi/2|)/2$. Hence, $S(\rho_B)\ug h[(1\piu s |\cos \varphi/2|)/2]$. 
{Using again \eqs(\ref{decomp}) and (\ref{coeffs}), the eigenvalues of $\rho$ are calculated as $(1\!\pm\! s)/4$, each being two-fold degenerate yielding that
 $S(\rho)\ug 1\piu h[(1\piu s)/2]$}. Using these results along with (\ref{inf}) in the light of (\ref{dis}), we find the QD of any state (\ref{class}) such that $\tau_0$ and $\tau_1$ have the same Bloch-vector length $s$ in the compact form
\begin{equation}\label{dis2}
\mathcal{D}\!=\!h\left[\tfrac{1\piu s |\cos \varphi/2|}{2}\right]\piu h\left[\tfrac{1\piu s |  \sin \varphi/2|}{2}\right]\meno h\left[\tfrac{1\piu s}{2}\right]
\meno 1.
\end{equation}
As for the classical correlations $\mathcal{C}\ug\mathcal{I}\meno\mathcal{D}$, using that $S(\rho_A)\ug S(\openone_A/2)\ug1$ we find that
\begin{equation}\label{cc}
\mathcal{C}\!=\!1\meno h\left[\tfrac{1\piu s |  \sin \varphi/2|}{2}\right].
\end{equation}
In \fig3 we plot $\mathcal{D}$ (a) and $\mathcal{C}$ (b) \vs $s$ and $\varphi$ as given by \eqs(\ref{dis2}) and (\ref{cc}). For a given $s$, \ie for fixed purity, $\mathcal{D}$ solely depends on $|\varphi\meno \pi/2|$ exhibiting its maximum value when $\varphi\ug \pi/2$. As $|\varphi\meno \pi/2|$ grows the QD progressively decreases until it vanishes at $\varphi\ug0,\pi$. On the other hand, for a  given  $\varphi$,  $\mathcal{D}$ increases with the purity at a rate that grows as $\varphi$ approaches $\pi/2$. Within the present class of states, the QD thus takes its maximum value  $\mathcal{D}_{\rm max}\!\simeq\!0.202$ [about three times larger than the maximum value attained in \fig1(a)] for $s\ug1$ and $\varphi\ug \pi/2$, which correspond to the B92 resource state $\rho_{{\rm B92}}$ introduced previously. This feature is in accordance with Ref.~\cite{gerardo} (where a different QCs' measure was used).
As for $\mathcal{C}$, \fig3(b) shows that this is maximum for $s\ug1$ and $\varphi\ug\pi$ and decreases when either of such two parameters is reduced.

We are now in a position to provide a comprehensive explanation for the effect in \fig1. Using the above discussed Kraus decomposition of $\mathcal{E}_p$ it is straightforwardly found that $\mathcal{E}_p(|\pm\rangle\langle\pm|)\ug(\openone\!\pm\!\sqrt{1\meno p}\,\sigma_1\piu p\sigma_3)/2$. Hence, $s_{0x}\ug-s_{1x}\ug \sqrt{1\meno p}$, while $s_{0z}\ug s_{1z}\ug p$ [note that this immediately shows that $s_0\!\equiv\! s_1\ug s$ for any $p$]. Using this, we end up with
 \begin{equation}\label{Pphi}
s(p)\ug \sqrt{1\piu p(p\meno1)}\,,\,\,\,\,\,\varphi(p)\ug \pi\meno2\arctan\left({p}/{\sqrt{1\meno p}}\right)\,\,.
\end{equation}
By replacing (\ref{Pphi}) in \eqs(\ref{dis2}) and (\ref{cc}), the functions
{$\mathcal{D}(p)$ and $\mathcal{C}(p)$ so obtained  reproduce the plots in \fig1.}

To shed light on the physical mechanism behind the effect, in \fig4 we plot on the $XZ$ plane of the Bloch sphere the parabolic trajectories of the Bloch vectors corresponding to $\mathcal{E}_p(|\pm\rangle\langle\pm|)$.
\begin{figure}
 \includegraphics[width=0.55\textwidth]{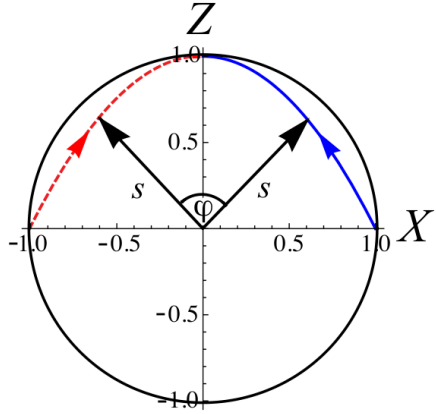}
\caption{(Color online) Trajectories of $\mathcal{E}_p(|+\rangle\langle+|)$ (blue solid line) and $\mathcal{E}_p(|-\rangle\langle-|)$ (red dashed) on the $XZ$ plane of the Bloch sphere. \label{Fig4}}
\end{figure}
Initially, the two vectors have unit length pointing towards opposite directions of the $X$-axis. Hence, $s\ug1$ and $\varphi\ug\pi$, which yields that $\mathcal{D}\ug0$ and $\mathcal{C}\ug1$ [\cf \eqs(\ref{dis2}) and (\ref{cc}) and \fig3]. When $p\ug1$, \ie for $\gamma t\!\gg\!1$, either state is mapped onto $|0\rangle\langle0|$, \ie the North Pole, which gives that $s\ug1$, $\varphi\ug0$ and thereby $\mathcal{D}\ug\mathcal{C}\ug0$ [\eq(\ref{class}) indeed shows that the state becomes evidently uncorrelated]. At the intermediate stage, map $\mathcal{E}_p$ acts in a way that the two Bloch vectors move along symmetrical parabolic trajectories (see \fig4) yielding that, in particular, $0\!<\!\varphi\!<\!\pi$. On the other hand, the purity shows a dip with $s$ taking its minimum $s_{\rm min}\ug\sqrt{3}/2$ for $p\ug1/2$ in accordance with (\ref{Pphi}).
In the light of \eq(\ref{dis2}) and \fig3(a), this {\it necessarily} brings about that $\mathcal{D}\!>\!0$ since $0\!<\!\varphi\!<\!\pi$ and the Bloch vector length, although shrinking to some extent, keeps finite throughout. This clarifies why discord must necessarily be created in the course of the process. In essence, the local dissipative channel transforms $|+\rangle\langle+|$ and $|-\rangle\langle-|$ so as to reduce their distinguishability, which unavoidably gives rise to QCs without entanglement \cite{seminal, talmor}. Note that the process non-unitarity is crucial since distinguishability as measured by the scalar product is unaffected by any unitary \cite{modi, nota_misura}. {Also, note that based on the definition of classically correlated states \cite{talmor} the above reasoning shows that QCs will be surely created regardless of the specific measure chosen to quantify them}.

In summary, in this Letter we asked whether QCs can develop as a result of local non-unitary dynamics, an unattainable phenomenon with entanglement. After deriving the QD for the class of involved states, we analytically proved that this indeed can occur for two qubits, initially in a fully classical state, under a local memoryless amplitude-damping channel. Also, we showed that the mechanism behind the {QCs'} birth can be readily grasped in the Bloch-sphere picture. 

{All these phenomena are arguably not restricted to qubits. The generalization to continuous-variable systems in a way that local bosonic Gaussian maps now play the role of the amplitude-damping channel is under ongoing investigations.}

{We thank R. Fazio and M. Paternostro for comments, and acknowledge support from FIRB IDEAS through project RBID08B3FM.}

{\it Note added.-}During completion of this work we became aware of a related manuscript \cite{campbell}.

\end{document}